\documentclass[twocolumn,showpacs,preprintnumbers,amsmath,amssymb]{revtex4}

\usepackage{graphicx}
\usepackage{dcolumn}
\usepackage{bm}
\usepackage{amsmath}

\usepackage{color}

\begin{document}


\title{Absolute frequency determination of molecular transition in the Doppler regime at kHz level of accuracy}
\author{K. Bielska, S. W\'{o}jtewicz, P. Morzy\'{n}ski, P. Ablewski, A. Cygan, M. Bober, J. Domys{\l}awska, M. Zawada, R. Ciury{\l}o, P. Mas{\l}owski and D. Lisak}
\affiliation{Institute of Physics, Faculty of Physics, Astronomy and Informatics, Nicolaus Copernicus University in Toru\'{n}, Grudziadzka 5, 87-100 Torun, Poland}

\begin{abstract}
We measured absolute frequency of the unperturbed P7 P7 O$_2$ B-band transition $\nu_0=$~434783.5084857(82)~GHz and the collisional self-shift coefficient $\delta=-9.381(62)\times10^{-21}$~GHz/(molecule/cm$^3$). With Doppler-limited spectroscopy we achieved the relative standard uncertainty of $2\times10^{-11}$ on line position, typical for Doppler-free techniques. Shapes of the  spectral line were measured with a Pound-Drever-Hall-locked frequency-stabilized cavity ring-down spectrometer referenced to an $^{88}$Sr optical atomic clock via an optical frequency comb.   
\end{abstract}


\maketitle

\section{Introduction}

High precision and accuracy in measurement is required for searching for subtle physical effects and small discrepancies between the experiment and theory \cite{Woo97,Hud11,Sal13}. 
Spectroscopy of weak molecular transitions, like the O$_2$ B-band \cite{Wci13} line presented here or the quadrupole transitions in H$_2$ or D$_2$ \cite{Mon16}, can provide valuable information about molecular structure \cite{Yu12} or verify quantum electrodynamics in molecular systems \cite{Kom11}.  
Unfortunately, Doppler-free techniques in gas cells or even in optical cavities cannot be applied for such transitions.
Therefore, Doppler-limited spectroscopy techniques are being continuously developed to reach the accuracy of molecular transitions frequencies determination at the level typical for Doppler-free techniques \cite{Bur15,Gat16}.

Here we present the results of determination of the unperturbed frequency and collisional shift of the P7 P7 line in O$_2$ B band. Our uncertainty on the transition frequency is about 20 times lower than in the case of previous results reported for this band \cite{Dom16}. 
The experimental spectra were collected here with an optical frequency comb (OFC) assisted, Pound-Drever-Hall locked  frequency-stabilized cavity ring-down spectrometer (PDH-locked FS-CRDS) \cite{Cyg11_3}. 
An $^{88}$Sr optical atomic lattice clock \cite{Bob15,Morz15} served as a frequency reference in most of the performed measurements. The spectrometer was optically linked to the optical atomic clock via the OFC \cite{Did00,Udem02}. Such a reference has significant advantages over typically used radio frequency (RF) references in terms of stability and accuracy. 
Optical atomic clocks \cite{Lud15} are well recognized as a state of the art technology for time and frequency metrology.
The ultra-narrow optical atomic transition frequencies measured with optical atomic clocks are recommended by the BIPM (Bureau International des Poids et Mesures) for practical realizations of the metre and the second \cite{SI}. 
The approach described here is a proof-of-principle experiment exploiting an optical atomic clock as a frequency reference in molecular spectroscopy.

Our relative combined uncertainty of $2\times10^{-11}$ for the unperturbed frequency of molecular transition is one of the lowest values obtained in Doppler-limited molecular spectroscopy \cite{Tru13,Gat15}.

\section{Experimental apparatus}

\begin{figure}
	\includegraphics[width=8cm]{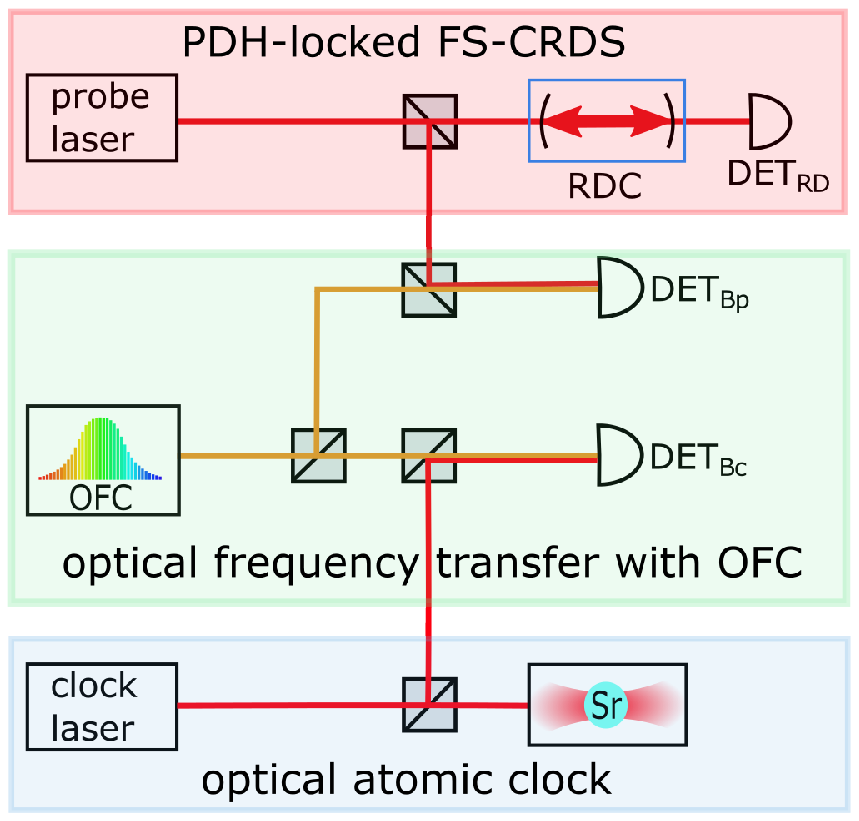}
	\caption{(Color online) Schematic diagram showing links between main parts of the experimental setup. RDC -- ring-down cavity, DET$_{\rm RD}$, DET$_{\rm Bp}$ and DET$_{\rm Bc}$ --  photodiodes.}
	\label{fig:setup}
\end{figure}

The experimental setup consists of three subsystems, see Fig. \ref{fig:setup}. The first is the PDH-locked FS-CRDS used for probing the P7 P7 transition \cite{Cyg11_2,Cyg11_3}. The second one is the optical atomic clock which provides the optical frequency reference \cite{Bob15,Morz15}. The third one is the optical frequency comb, which transfers optical frequency between other subsystems \cite{Ros08}. The OFC is also used to determine the absolute frequency if measurements are not referenced to the optical atomic clock but to another frequency reference.

\subsection{Spectrometer}
The heart of the PDH-locked FS-CRDS \cite{Cyg11_3} is a 74 cm long, high-finesse optical cavity \cite{Hod04}. This length corresponds to the free spectral range (FSR) of about 204 MHz. The cavity mirrors are spherical (radius of curvature $r=1\ {\rm m}$) and dual-wavelength coated. Their nominal reflectivity is $R=0.99975$ in the spectral range near $\lambda = 690 \textrm{ nm}$ and $R=0.98$ for $\lambda = 633 \textrm{ nm}$, corresponding to the O$_2$ B band and to the wavelength of a helium-neon laser (HeNe), respectively. 
One of the cavity mirrors is mounted on a piezo transducer, which is used for active stabilization of the cavity length to multiples of half wavelength of light emitted by the frequency stabilized HeNe laser. 
The probe laser is an external cavity diode laser (ECDL) spectrally narrowed and actively locked to the cavity resonance by the PDH technique with active correction of the PDH error signal offset \cite{Cyg11}. Our spectra were measured with a step size of about 50 MHz, which was achieved by controlled changes of the cavity length, modifying the frequency of the HeNe laser with an acousto-optic modulator (AOM) working in double pass configuration \cite{Hod04}. The absorption coefficient of the sample is determined from the decay time constant of light leaking from the cavity after the probe laser is turned off. Further details of the spectrometer's operation can be found in Ref. \cite{Dom16} and references therein.

\subsection{Absolute frequency determination}
We have used two absolute frequency references. The first is a 10 MHz RF signal of a hydrogen maser \cite{Jiang15}. This signal serves as a link to the UTC(AOS) and is delivered to our laboratory by a 330~km long, frequency-stabilized fibre link \cite{Morz15,Kre15} from Astro-Geodynamic Observatory (AOS) in Borowiec. All frequency counters and generators used in the PDH-locked FS-CRDS as well as in the OFC are referenced to this signal \cite{Cyg16}. The other one is provided by the $^{88}$Sr optical lattice clock in the National Laboratory FAMO in Toru\'{n}. They enable two schemes of absolute frequency determination in measured spectra, where the first one incorporates only microwave frequency standard and the second one uses in addition the optical clock.

To realize the optical frequency reference we use a 698 nm clock laser: an ECDL tightly locked to a mode of a high-finessee optical cavity, providing a short term (1 s) relative instability of the laser frequency below $5\times10^{-14}$. The clock laser shifted by an AOM is digitally locked \cite{Morz13} to the ${^1}S_0-{^3}P_0$ clock transition in ultracold $^{88}$Sr atoms confined in a 1D optical lattice \cite{Bob15,Morz15}. 
Long term frequency instability is below $10^{-15}$ after 3 hours of averaging. 
The absolute frequency $f_c$ of the $^1S_0$ -- $^3P_0$ transition in bosonic $^{88}$Sr
referenced to the Coordinated Universal Time (UTC) is equal to 429~228~066~418~008.3(1.9)$_{\rm syst}(0.9)_{\rm stat}\ {\rm Hz}$ \cite{Morz15}. 

The Er:fibre OFC is used to bridge the frequency span between the spectrometer and the optical atomic clock, as well as the RF frequency reference. The OFC
spectrum is frequency shifted  in a commercial  non-linear amplifier to the range around 1390 nm and subsequently  frequency doubled in an MgO:PPLN crystal. The resulting spectrum covers simultaneously the spectrometer and clock laser wavelengths: 690 nm and 698 nm, respectively.
The frequency of $n$-th comb tooth in the frequency doubled OFC is given by 
\begin{equation}
f_n=2f_0+n f_{\rm rep}
\label{f_comb}
\end{equation} 
where $f_0$ is the OFC offset frequency and $f_{\rm rep}$ is its repetition rate. 
For each frequency of the spectrometer probe laser, two heterodyne beat-note frequencies are measured simultaneously:  $f_{\rm Bc}$ between the OFC and the clock laser, and $f_{\rm Bp}$ between the OFC and the spectrometer probe laser. 
The frequencies $f_0$ and $f_{\rm rep}$ remain unchanged (locked to the hydrogen maser) during entire spectrum acquisition, whereas the variable spectrometer probe beam frequency is additionally shifted with an AOM to keep $f_{\rm Bp}$ in the convenient for measurement range near 60~MHz. The spectrometer probe laser frequency can be determined as
\begin{equation}
f=f_c+\left(n_p-n_c\right)f_{\rm rep}-f_{\rm Bc}+f_{\rm Bp}-f_s=f_c+\Delta f,
\end{equation}
where $n_p$ and $n_c$ denote comb tooth numbers for the beat-note with spectrometer probe and clock lasers, respectively, and they both are close to $1.7\times10^6$. A sum of all additional controlled frequency shifts introduced in both systems by acousto-optic modulators is denoted as $f_s$. 
The total uncertainty of $f$ comes from the optical reference $f_c\approx4.3\times10^{14}$~Hz having $5\times10^{-14}$ relative uncertainty at 1 s  and RF reference having 10$^{-12}$ relative uncertainty at 1 s.
For $n_p-n_c\approx2.2\times10^4$ and $f_{\rm rep}\approx2.5\times10^8$~Hz, the RF reference contribution to the relative uncertainty of $f$ is $1.3\times10^{-14}$ at 1 s. In this scheme only the frequency difference $\Delta f$ between $f$ and $f_c$, two orders of magnitude smaller than $f$ itself, is affected by the instability of the RF reference. Therefore both components, the optical and RF references, lead to a total relative uncertainty of $f$ at a level typical for optical atomic clocks.

We have shown above that even though the OFC is not locked to the clock laser, but is self-referenced and stabilized to the RF signals, the use of an optical atomic clock as a frequency reference may significantly improve frequency measurement.
The frequency $f$ is determined as:
\begin{equation}
f=2f_0+n_pf_{\rm rep}+f_{\rm Bp}+f_s'.
\end{equation}
In that case $f_s'$ is again the sum of all the additional controlled frequency shifts introduced by the acousto-optic modulators in the PDH-locked FS-CRDS system. The relative uncertainty of $f$ would be the same as for the RF reference, that means $10^{-12}$ at 1 s in that case \cite{Cyg16}. This is two orders of magnitude worse than in the case of use of the optical atomic clock as a frequency reference.

\section{Measurements and data analysis}

\begin{figure}
	\includegraphics[width=8cm]{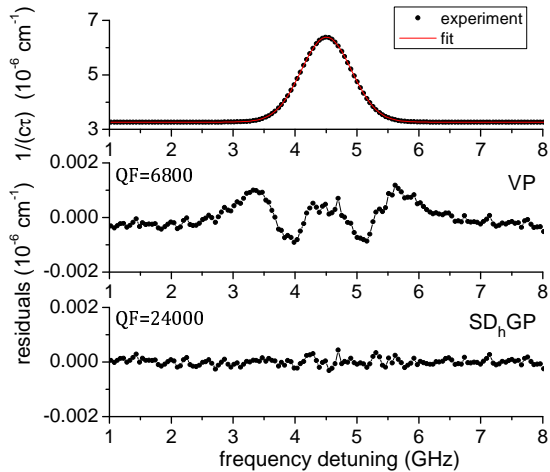}
	\caption{(Color online) The P7 P7 line measured at room temperature and pressure of 800 Pa (top), residuals from the VP (middle) and the SD$\rm _h$GP (bottom) individual fits.}
	\label{fig:residua}
\end{figure}

A high-purity oxygen sample (99.999\%) of natural isotopic abundance was used.
The P7 P7 line was measured at room temperature and four pressures between 200 and 800 Pa, which corresponds to number density $N$ range from $0.49\times10^{17}$ to $1.9\times10^{17}\ {\rm molecules/cm^3}$. Data for three of those pressures were acquired within one week and referenced to the optical atomic clock. The fourth pressure ($\sim$ 600 Pa) was measured four months later with the hydrogen maser being the only frequency reference. Both sets of data are in excellent agreement, which indicates a high repeatability of the measurement. 
Multiple spectra were acquired and averaged at each pressure.  The number of recorded spectra varied between 40 and over 100 per pressure. 
Several line-shape models were used in the data analysis, all of which include the Doppler broadening, the collisional broadening and shifting. We also considered the velocity changing collisions in terms of the soft-collision model (the Galatry profile, GP) \cite{Gal61} and the hard-collision model (the Nelkin-Ghatak profile, NGP) \cite{Nel64}, as well as their speed-dependent versions: the speed-dependent Galatry profile (SDGP) \cite{Ciu97} and the speed-dependent Nelkin-Ghatak profile (SDNGP) \cite{Lan97}, respectively. We additionally tested the speed-dependent Voigt profile (SDVP) \cite{Ber72}, which includes the speed dependence of collisional broadening and shifting but neglects the velocity changing collisions. Speed dependence of the line-shape parameters was described either in terms of quadratic approximation \cite{Pri00} or with confluent hypergeometric functions \cite{War74}, which we denote in the profile name abbreviation by subscripts 'q' and 'h', respectively. Unperturbed line positions and collisional shift coefficients derived with all above models were compared to the values extracted by fitting the simple Voigt profile (VP). We found that the P7 P7 line shows no detectable asymmetry, thus the line positions and collisional shift coefficients retrieved with each of the fitted profiles agree with each other and the VP results to one standard uncertainty. Fit residuals for the highest sample pressure, for the SD$\rm _h$GP profile and for the VP are shown in Fig. \ref{fig:residua}. Fits of spectra with those profiles resulted in the highest and lowest quality of the fit (QF) \cite{Cyg12}, respectively.

\begin{figure}
	\includegraphics[width=8cm]{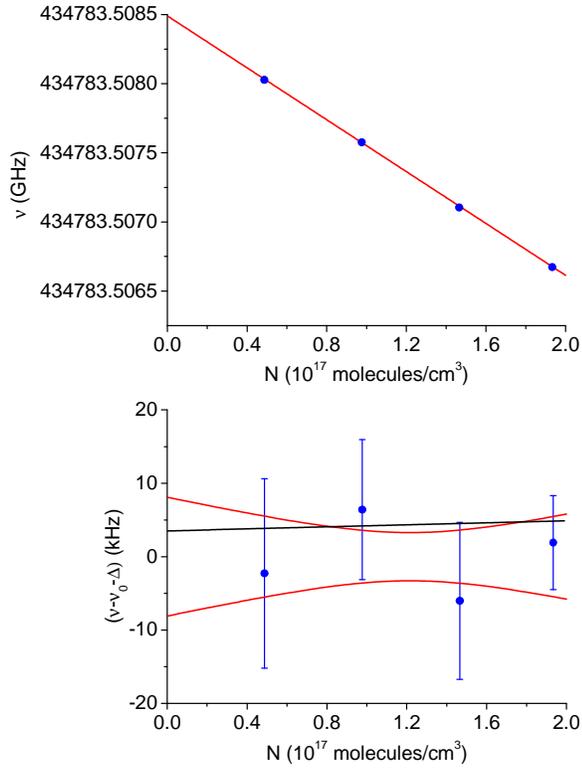}
	\caption{\label{fig:niepewnosci}(Color online) Top panel: P7 P7 line center frequencies $\nu$ vs absorber concentration $N$ obtained from fits of the SD$\rm _h$GP profile for each pressure separately (blue points) and least-square linear fit (red line). Bottom panel: least-square fit uncertainties extrapolated to zero concentration (red lines) and line positions shifted by determined value of unperturbed line position $\nu_0$ and collisional shift $\Delta$ (blue points) together with their uncertainties form individual fits of the SD$\rm _h$GP. Those uncertainties are about 20 times smaller than in our previous work \cite{Woj14}. The black line indicates the difference between line position calculated from parameters obtained in multispectrum fitting approach and when spectra for each pressure were fitted separately.}
\end{figure}

The data analysis was performed in two different approaches. In the first approach we fitted spectra for each pressure separately, while in the other one we applied a multispectrum fitting technique \cite{Ben95,Pin01}. 
The results obtained from a multispectrum fit on unperturbed line position is slightly higher, and on collisional shift is slightly lower than from individual fits. These differences are the same for each profile used. Nevertheless, the values agree within standard uncertainty.
The dominant component in the total uncertainty is type A, mostly related to the instability of the ring-down cavity caused by the reference HeNe laser frequency jitter. Type B uncertainty is associated with pressure and temperature determination, see Table \ref{tab:uncert}. 
We have also tested the influence of the probe laser power on the determined line position \cite{Cza04}. 
We performed a set of four additional line position measurements at pressure of 400~Pa and 600~Pa, and the intra cavity power range from 0.05~W up to 0.45~W which exceeds power variations in the main set of data shown in Fig. \ref{fig:niepewnosci}.
We conclude that in our experimental conditions the power shift of the line is below the detection limit.

The absolute line positions $\nu$, extracted for different sample concentrations $N$, are shown in the top panel of Fig. \ref{fig:niepewnosci}. From these values we determined the collisional shift coefficient $\delta=\Delta/N$ and the  unperturbed line position $\nu_0$ obtained from the extrapolation of $\nu$ to zero concentration. Here $\Delta$ is the  collisional shift of the spectral line proportional to the gas concentration. The uncertainties and residuals for a linear least-square fit to measured line positions are shown in the bottom panel of Fig. \ref{fig:niepewnosci}. The measured line positions agree with a linear fit on a concentration of absorber within their uncertainties.
Based on the data analysis performed with the SD$\rm _h$GP we obtained the value of the unperturbed transition frequency equal to $\nu_0=$~434783.5084857(82)~GHz and the collisional shift coefficient equal to $\delta=-9.381(62)\times10^{-21}$~GHz/(molecule/cm$^3$). 
These values agree with our previous results within the combined standard uncertainty for line position: $\nu_0=$~434783.50841(17)~GHz and within three standard deviations for the collisional shift coefficient: $\delta=-8.90(16)\times10^{-21}$~GHz/(molecule/cm$^3$) determined with the SD$\rm _q$VP \cite{Woj14}. 
Our measurement uncertainties correspond to relative uncertainties of $2\times10^{-11}$ for the transition frequency and $7\times10^{-3}$ for the collisional shift coefficient.

\section{Conclusions}

We have demonstrated a Doppler-limited experimental technique capable of precise determinations of unperturbed frequencies of molecular transitions. We obtained the most accurate absolute transition frequency in the O$_2$ B band. The uncertainty of the line position is 20 times lower than the previous most accurate result \cite{Woj14}. Measurements of even several ultra-accurate oxygen line positions are of great importance for improvement of molecular constants and the global model of O$_2$ spectra in a broad spectral range \cite{Yu12}.

This proof-of-principle experiment introduces optical atomic clocks as a frequency reference to Doppler-limited molecular spectroscopy. The approach presented here will allow to push the relative accuracy on transition frequency determination below $10^{-12}$ if cavity stability and signal-to-noise ratio of measured spectra is improved.

The proposed technique can be used for accurate measurements of line positions and collisional shift coefficients for lines which are crucial for fundamental science and cannot be easily accessed by Doppler-free techniques, such as weak hydrogen lines \cite{Mon16}. Frequency measurements of these transitions find applications in verification of quantum electrodynamics in molecular systems for which high-accuracy {\it ab initio} calculations are possible \cite{Kas12}. Moreover, this kind of study can set upper bounds on possible fifth-force interactions, which would indicate physics beyond the Standard Model \cite{Sal13}.

\begin{acknowledgments}
The research is part of the program of the National Laboratory FAMO in Toru\'{n}, Poland. The reseach is supported by the National Science Center, Poland project numbers 2014/15/D/ST2/05281, DEC-2012/05/D/ST2/01914, DEC-2013/11/D/ST2/02663, 2015/17/B/ST2/02115 and by the COST Action, CM1405 MOLIM. It has also received funding from EMPIR programme co-financed by the Participating States and from the European Union's Horizon 2020 research and innovation programme (EMPIR 15SIB03 OC18). 
\end{acknowledgments}

\begin{table*}
	\centering
	\caption{\label{tab:uncert}Uncertainty budget and uncertainty type for each source of uncertainty for unperturbed line position $\nu_0$ and collisional shift coefficient $\delta$.}
	\begin{tabular}{lccc}
		\hline
		Source of uncertainty&Type&$u\left(\nu_0\right)$ (kHz) & $u\left( \delta \right) / \delta $ (\%) \\ 
		\hline
		Fit uncertainty&A& 8.1 &0.65\\
		Pressure offset (max 0.13 Pa)&B&0.18&---\\
		Pressure reading \& calibration (max 0.75 Pa)&B&0.60&0.1\\
		Temperature: variations, gradient, reading \& calibration (max 240 mK)&B&0.86&0.094\\
		\hline
		Combined uncertainty&&8.2&0.7\\	
		\hline
	\end{tabular}
\end{table*}

\end{document}